# Weak Interactions, Tunneling Racemization and Chiral Stability

M.Cattani[a] and J.M.F.Bassalo[b]

[a]Instituto de Física da Universidade de São Paulo, C.P.66318, 05315-970, SP, Brazil.
[b]Fundação Minerva, Av.Gov.José Malcher 629,.66035-10, Belém, PA, Brasil.
mcattani@ifusp.br  and  jmfbassalo@gmail.com

Abstract
   We study in the framework of the Schrodinger equation the effect of intermolecular interactions on the tunneling racemization of the active molecule. The active molecule is assumed as a two-level system and the left-right isomerism is viewed in terms of a double-bottomed harmonic potential well. The active molecule is assumed to be embedded in a gas, liquid or solid, submitted to a perturbing potential U created by the molecules of the sample. In our model we take into account the difference of energy E due to the weak interactions between the left (L) and right (R) configurations. We have shown that when E is equal to zero the system cannot be *optically stable*: the optical activity tends asymptotically to zero in the case of dilute gases or compressed gases and liquids or oscillates periodically around zero when the molecules are isolated or submitted to a static potential. Only when E is different of zero the system can be *optically stable* depending on the strength parameters of the potential U and on the magnitude of the spontaneous tunneling.

1. **Introduction**

   Today we know that chirality is one of the most exciting phenomena in nature as well in science.[1–4] From elementary particles to human beings, nature is not symmetry with respect to chirality, or L− and R− handedness. This fundamental aspect that was pointed out by Pasteur in 1857 conjecturing that "L´Univers est dissymétrique" was only confirmed in the middle of the 20$^{th}$ century.[1–4]

   Optical activity occurs[4–6] when the molecule has two distinct left and right configurations, │L > and │R >, which are degenerate for a parity operations, i.e., $P(x)│L > = │R >$ and $P│R > = │L >$. Left − right isomerism can be viewed in terms of a double-bottomed potential well (see Fig.1) and the states │L > and │R > may be pictured as molecular configurations that are concentrated in the left or right potential well. The two enantiomers of a chiral molecule are described by the superposition of the odd and even parity eigenstates of the double well localized around the potential minima, x = a and x = − a. The coordinate x is involved in the parity operation P = P(x) and connects the two potential minima. It may represent the position of an atom, the rotation of a group around a bond, some other coordinate, or a collective coordinate of the molecule

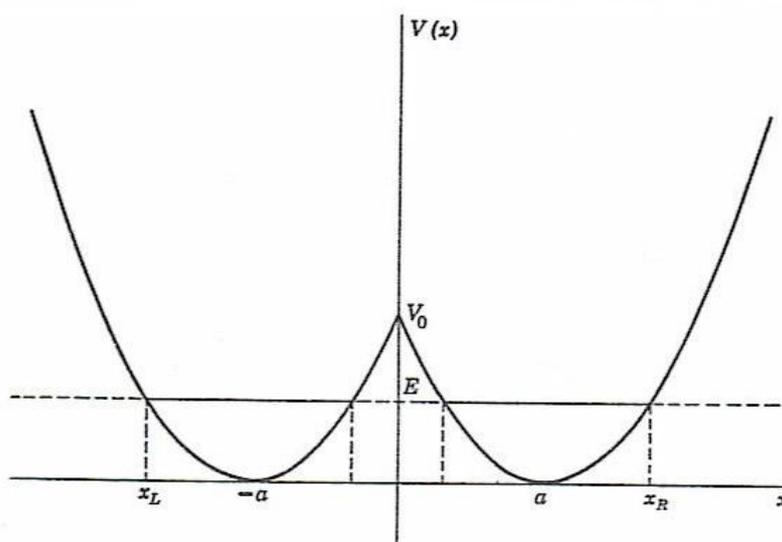

Figure 1. Double harmonic potential $V(x) = k(|x| - a)^2$; $x = -a$ and $x = a$ are the two points of minima. $V_o$ is the potential barrier separating the two minima of the double well. If $E < V_o$ there are two internal classical turning points $x_L$ and $x_R$, but the quantum tunnel effect permits penetration of the barrier.

As is well known[5,6], the optical activity of an optically active material changes with time. The sample, containing predominantly one stereoisomer, will become a mixture of equal amounts of each isomer. This relaxation process, which is called racemization, occurs spontaneously or is due to the interaction of the active molecule with the environment. Many approaches have been proposed to describe the racemization.[7] However, these models are not completely satisfactory because they involve some phenomenological parameters whose identification and quantification are not immediate.[2-7] The understanding of the racemization and stability of the optical activity is extremely important for the fundamental physical point of view[1-7] and for their practical applications in chemistry.[8-11] More and more modern drug design addresses the fact that enantiomers can have dramatic difference in their physical and pharmacological properties.[8]

Let us define by $H_o$ the Hamiltonian of each side of the double well and by $V_o$ the potential barrier separating the two minima of the double well. In this picture, $|L>$ and $|R>$ are eigenstates of $H_o$, i.e., $<L|H_o|L> =$ $< R \mid H_o \mid R > = E_o$ and there is a small overlap of these states inside the barrier V(x) so that , $< L \mid V \mid R > = < R \mid V \mid L > = \delta$.

Let us assume that the double-bottomed potential well has the shape of two overlapping harmonic potentials. Indicating by ω the fundamental frequency of each harmonic oscillator and by μ the reduced mass of the particles vibrating between x = a and x = − a, the fundamental vibrational states $\mid \Phi(x) >$ of the left and right harmonic oscillators are written, respectively, as:[12]

$$\mid \Phi_L(x) > = (\mu\omega/\pi\hbar)^{1/4} \exp[-(\mu\omega/2\hbar)(x + a)^2],$$

(1.1)

$$\mid \Phi_R(x) > = (\mu\omega/\pi\hbar)^{1/4} \exp[-(\mu\omega/2\hbar)(x - a)^2].$$

The L and R configurations states of the active molecule will be written in a Born-Oppenheimer approximation (adiabatic approximation) as $\mid L > = \mid \psi_L > \mid \Phi_L(x) >$ and $\mid R > = \mid \psi_R > \mid \Phi_R(x) >$, where $\mid \psi >$ describes all internal degrees of freedom of the active molecule except x .

So, the parameter responsible for the *spontaneous* or *natural tunneling* between the L and R configurations, defined by $\delta = < L \mid V(x) \mid R > = < R \mid V(x) \mid L >$ is given by:[12]

$$\delta = (h\omega/\pi^{3/2})(\mu \omega a^2/\hbar)^{1/2} \exp(-\mu \omega a^2/\hbar). \quad (1.2)$$

A good numerical estimation of δ/h can be obtained taking into account typical molecular parameters: $a = 10^{-8}$ cm and $\mu = 10^{-23}$ g. Writing ω as ω = A $10^{13}$ rad/s and using Eq.(1.2) the parameter δ/h., measured in Hz or in years, is given by

$$\delta/h = 5.54 \; 10^{12} \; A^{3/2} \exp(-9.52 \; A) \; Hz = 1.75 \; 10^{20} \; A^{3/2} \exp(-9.52 \; A) \; y^{-1} \quad (1.3).$$

The spontaneous oscillation period τ between the L and R configurations will be indicated by τ = h/δ. For frequencies in the infrared region, in the range $4.8 \; 10^{13} \leq \omega \leq 5.8 \; 10^{13}$ rad/s, the period τ varies in the interval 15 days ≤ τ ≤ 390 y, respectively.

Many optical experiments[13,14] have demonstrated cases in which mirror symmetry in stable atoms is broken during absorption of light. These results support the theory of unification of the electromagnetic and weak forces.[1,3] The discovery of parity violation in an atomic process was the outcome of many years of experimental effort. After the emergence of unified theories in the early 1970´s, many experiments were designed to test the new theories, to choose between them and to measure the fundamental constants involved.[14]

If weak interaction effects are present, parity is violated and the L and R sides of the double-bottomed potential are no longer symmetrical. In this way, $< L | H_o | L > = E_L = E_o - \varepsilon$ and $< R | H_o | R > = E_R = E_o + \varepsilon$, where $2\varepsilon$ is the difference of energy between the L and R configurations due to the parity −violating interaction. According to recent calculations,[15−21] $\varepsilon/h$ is typically of the order of $10^{-3}$ Hz for rotational and vibrational transitions and of the order of $10^{-6}$ Hz for nuclear magnetic transitions.

In Section 2 we write the Schrödinger equation to calculate the transitions between the states L and R in the general case, that is, taking into account simultaneously: (a) the spontaneous tunneling effect, (b) the energy difference $\varepsilon$ due to the weak forces and (c) when the active molecule is submitted to a generic perturbing potential U(t). In Section 3 and 4 using the general expressions deduced in Section 2 the *racemization* and the *optical activity* are calculated when $\varepsilon = 0$ and $\varepsilon \neq 0$ for some different potentials U(t). It will be shown that depending on the $\varepsilon$ value and on U(t) the active sample can become optically stable. In Section 5 we present the Summary and Conclusions and, finally, in Section 6 we present the Discussions.

## 2. Active molecule interacting with the environment.

In precedent papers[4,22−3] we have calculated, using the Schrödinger`s equation, the racemization when the active molecules, embedded in a gas, liquid or solid, is submitted to a generic U(t). In our approach we have assumed that the racemization is produced essentially by transitions between the two vibrational states $|L>$ and $|R>$. In this way, the state function $|\Psi(t)>$ of the active molecule, is represented by

$$|\Psi(t)> = a_L(t)|L> + a_R(t)|R>, \quad (2.1)$$

and obey the equation $i\hbar \, \partial|\Psi(t)>/\partial t = [H_o + V(x) + U(t)]|\Psi(t)>$. So, $a_L(t)$ and $a_R(t)$ are governed by the following differential equations:

$$da_L(t)/dt = -(i/\hbar)[a_L(t)(E_o - \varepsilon + U_{LL}) + a_R(t)(\delta + U_{LR})],$$
$$(2.2)$$
$$da_R(t)/dt = -(i/\hbar)[a_R(t)(E_o + \varepsilon + U_{RR}) + a_L(t)(\delta + U_{RL})],$$

where the matrix elements $U_{nk}$, with n,k = L and R, are given by $U_{nk}(t) = <n|U(t)|k>$.

Since the homochiral interactions are equal,[31] we define $u(t) = U_{LL}(t) = U_{RR}(t)$. The heterochiral interaction will be indicated by $\varphi(t) = U_{LR}(t) = U_{RL}(t)$. In this way Eqs.(2.2) are written as:

$$da_L(t)/dt = -(i/\hbar)[a_L(t)(E_o - \varepsilon + u(t)) + a_R(t)(\delta + \varphi(t))],$$

$$da_R(t)/dt = -(i/\hbar)[a_R(t)(E_o + \varepsilon + u(t)) + a_L(t)(\delta + \varphi(t))].$$

(2.3)

In next Sections the general Eqs.(2.3) will be adopted to calculate the *racemization* and the *optical activity* or *chiral activity* for some particular conditions. In Section 3 we assume that $\varepsilon = 0$ and (1) $U = 0$, (2) $U =$ static $\neq 0$ and (3) $U = U(t)$ for dilute gases and for compressed gases and liquids. In Section 4 we analyze the cases when $\varepsilon \neq 0$ and (1) $U =$ static $\neq 0$ and (2) $U = U(t)$ for dilute gases and compressed gases and liquids

## (3) $\varepsilon = 0$.

Assuming that the weak interactions are negligible ($\varepsilon = 0$) we will study three different cases: when the active molecule is isolated ($U = 0$), when it is submitted to a perturbing static potential $U =$ static $\neq 0$ and when it is immersed in a dilute gas or in a compressed gas or liquid submitted to $U = U(t)$.

Putting $\varepsilon = 0$ in Eqs.(2.3), $a_L(t)$ and $a_R(t)$ obey the following equations:

$$da_L(t)/dt = -(i/\hbar)[a_L(t)(E_o + u) + a_R(t)(\delta + \varphi)],$$

$$da_R(t)/dt = -(i/\hbar)[a_R(t)(E_o + u) + a_L(t)(\delta + \varphi)].$$

(3.1)

These equations can be solved exactly giving:

$$a_L(t) = \exp[-i(E_o t/\hbar + \theta_{LL}(t))][a\exp(-i\theta_{LR}(t)) + b\exp(i\theta_{LR}(t)]/2,$$

$$a_R(t) = \exp[-i(E_o t/\hbar + \theta_{LL}(t))][a\exp(-i\theta_{LR}(t)) - b\exp(i\theta_{LR}(t)]/2,$$

(3.2)

where a and b are constants determined by the initial conditions and

$$\theta_{nk}(t) = \int_0^t <n|V(x) + U(t)|k> dt/\hbar,$$

with n, k = L and R.

If at t = 0 the active molecule is prepared so that $|\Psi(0)> = |L>$, we obtain from Eqs.(3.2), putting $a_L(0) = 1$ and $a_R(0) = 0$, a = b = 1. Therefore, the active molecule state will be described by:

$$|\Psi(t)> = \exp[-i(E_o t/\hbar + \theta_{LL}(t))] \{\cos[\theta_{LR}(t)] |L> - i \sin[\theta_{LR}(t)] |R>\} \quad (3.3)$$

In this way, the *racemization rate* or simply *racemization* r(t) will be given by,[22,24–26]

$$r(t) = \{|<R|\Psi(t)>|^2\} = \{\sin^2[\theta_{LR}(t)]\}, \quad (3.4)$$

where $\theta_{LR}(t) = \int_0^t <L|V(x) + U(t)|R> dt/\hbar = \delta t/\hbar + \int_0^t \varphi(t) dt/\hbar$

and the brackets {…} mean an average over all perturbing effects of the potential U(t).

The *optical activity* or *chiral activity* O(t) of the sample is defined[15] in terms of the *racemization* r(t) by the equation O(t) = 1−2r(t). As the active molecule at t = 0 was prepared at the L configuration, its initial optical activity is $O_L = + 1$. At the R configuration the optical activity will be $O_R = - 1$.

(3.1) *U = 0.*

Putting U = 0 = φ into Eq.(3.4) we get,

$$r(t) = \{|<R|\Psi(t)>|^2\} = \sin^2(\delta t/\hbar) \quad (3.5).$$

So, when ε = 0 and the active molecule is isolated, that is, U = φ = 0 we have $r(t) = \sin^2(\delta t/\hbar) = [1 - \cos(2\delta t/\hbar)]/2$. Consequently, the *optical activity* O(t) = 1−2r(t) would be given by

$$O(t) = \cos(2\delta t/\hbar) \quad (3.6),$$

showing that O(t) oscillates with a period T = h/2δ between + 1 and −1, around the average value 0. Note that T = τ/2, where τ is the *spontaneous* or *natural tunneling period* defined by Eq.(1.2). For frequencies in the infrared region, in the range $4.8 \times 10^{13} \leq \omega \leq 5.8 \times 10^{13}$ rad/s, τ varies in the interval 15 days $\leq \tau \leq$ 390 y, respectively. For example, for $\omega \geq 5.8 \times 10^{13}$ rad/s we have T ≥ 195 y, implying that for these ω values an isolated molecule can remain active for a long period of time.

**(3.2) $U \neq 0$ static**

Let us assume that the active molecule is embedded in a dense gas, liquid or solid, where multiple interactions dominate over binary interactions and that there is a cooperative effect between the interacting molecules. Due to this collective behavior we will assume that each molecule is subjected to a *mean field* resulting from these combined interactions of all other molecules in the system. This *mean field* is understood as a *self-consistent Hartree field*. This cooperative interaction potential is static and will be indicated by U(x). Consequently, putting U = U(x) into Eqs.(3.4) the *racemization* r(t) is given by

$$r(t) = \sin^2[(\delta + \varphi)t/\hbar] = \{1 - \cos[2(\delta + \varphi)t/\hbar]\}/2 \quad (3.7)$$

and, consequently, the *chiral activity* O(t) = 1 − 2r(t):

$$O(t) = \cos[2(\delta + \varphi)t/\hbar] \quad (3.8)$$

Eq.(3.8) shows that the *optical activity* is not stable: O(t) oscillates between +1 and −1, around the average value 0 with a period T given by T* = 1/[2(δ/h + φ/h)].

In the Appendix the function φ is estimated in the particular case of dense gases and liquids composed by dipolar molecules where a cooperative interaction mechanism appears between the molecules of the sample. According to Eq.(A.1) the potential φ is given by,

$$\varphi = (\theta d/R^4) \exp(-\mu\omega a^2/\hbar) \quad (3.9),$$

where θ and d are quadrupole and dipole moments, respectively, of the molecules of the sample and *R* is the average distance between the interacting molecules. The factor φ/ℏ can be numerically estimated taking into account typical molecular values θ = $10^{-26}$ esu, d = $10^{-18}$ esu, μ = $10^{-23}$ g, a = $10^{-8}$ cm, putting $R \approx 3 \cdot 10^{-8}$ cm and ω = A $10^{13}$ rad/s. In this conditions we obtain

$$\varphi/h = 1.51 \cdot 10^{12} \exp(-9.52 \, A) \text{ Hz} \quad (3.10),$$

showing, according to Eq.(1.3), that δ/h ≈ φ/h. Taking into account the δ/h estimates in Section (3.1) we see that for ω ≥ 5.8 $10^{13}$ rad/s the period T* defined by Eq.(3.8) is given by T* ≥ 97.5 y, implying that for these ω values a molecule submitted to a static potential U(x) can remain active for a long period of time.

### (3.3) $U=U(t)$ produced by binary, additive and independent random collisions.

In this section we assume that the active molecule is embedded in dilute gases and in a compressed gases or liquids where the potential U(t) between the molecules of the sample is due to binary, additive and independent random collisions.[30,32] The interaction potential U(t) between the molecules is described by a sum of binary interactions given simply by $\gamma/R(t)^p$ where $\gamma$ is the constant of force between the interacting particles, R(t) the distance between them as a function of the time t and the exponent p = 4,5,…, if the interaction is dipole–quadrupole, quadrupole–quadrupole,…

The molecular collisions which induce transitions between L and R configurations are described by $\varphi(t)$ in Eqs.(2.2). The spontaneous transitions between L and R are described by $\delta$. Putting $\varepsilon = 0$ into Eqs.(2.2), we verify that $a_L(t)$ and $a_R(t)$ obey the following equations:

$$da_L(t)/dt = -(i/\hbar)[a_L(t) (E_o - u(t)) + a_R(t) (\delta + \varphi(t))],$$
(3.11)
$$da_R(t)/dt = -(i/\hbar)[a_R(t) (E_o + u(t)) + a_L(t) (\delta + \varphi(t))].$$

In *dilute gases*, for molecular densities $N\sim 10^{17}/cm^3$ the collisions have very short duration (around $10^{-11}$s for a system at room temperature) and a very high collision frequency. Calculating the collisions effects using the *impact approximation* we have shown that[25,26] the *racemization* $r_1(t)$ is given by,

$$r_1(t) = [1 - \cos(2\delta t/\hbar) \exp(-\lambda t)]/2,$$
(3.12)

where $\lambda = (\gamma/\hbar)^{2/(p-1)} N(kT/m)^{(p-3)/(2p-2)}$, N the density of perturbing molecules, k the Boltzmann constant, T the absolute temperature of the system and m the reduced mass of the colliding particles. In this case the *optical activity* $O_1(t) = 1 - 2r_1(t)$ is given by

$$O_1(t) = \cos(2\delta t/\hbar) \exp(-\lambda t)$$
(3.13).

For *compressed gases or liquids*, where collisions are *quasi-static*, we have shown that[30] the *racemization* $r_2(t)$ is given by:

$$r_2(t) = [\,1 - \cos(2\delta t/\hbar)\exp(-\lambda^* t^{3/p})]/2 \,, \qquad (3.14)$$

where $\lambda^* = (8\pi/p)N(\gamma/2\hbar)^{3/p} \int_0^\infty x^{-(p+3)/p} \sin^2 x \, dx$. Consequently the optical ativity $O_2(t) = 1 - 2r_2(t)$ becomes

$$O_2(t) = \cos(2\delta t/\hbar)\exp(-\lambda^* t^{3/p}) \qquad (3.15).$$

Eqs.(3.13) and (3.15) show that for sufficiently large t values ($\lambda t \gg 1$ and $\lambda^* t^{3/p} \gg 1$) the *optical activities* of the samples tends to zero. Thus, if $\varepsilon = 0$, we verify that in dilute gases or in compressed gases and liquids the molecular interactions would inevitably produce a complete racemization of the sample. From Eqs.(3.13) and (3.15) we verify that in dilute gases, $O_1(t)$ decays in time as $\exp(-\lambda t)$ and in dense gases and liquids $O_2(t)$ decays more slowly, as $\exp(-\lambda^* t^{3/p})$, since p = 4,5,…and so on.

In order to get simple estimates for $O_1(t)$ and $O_2(t)$ let us assume that there is only a dipole−quadrupole (p=4) interaction between active and perturbing molecules. In this case,[23] since $\gamma = d < L | Q(x) | R > = d\,\theta \exp(-\mu\omega a^2/\hbar)$, $\lambda$ given by Eq.(3.12) becomes :

$$\lambda = 13.0\, N\, (kT/m)^{1/6}\, (\theta d/\hbar)^{2/3}\, \exp(-2\mu\omega a^2/3\hbar)\,, \qquad (3.16)$$

where d is the electric dipole of the perturbing molecule and $\theta$ quadrupole matrix element of the active molecule between L and R configurations. Similarly,[30] $\lambda^*$ given by Eq.(3.14) becomes:

$$\lambda^* = 2.86\pi\, N(\theta d/2\hbar)^{3/4}\, \exp(-3\mu\omega a^2/4\hbar). \qquad (3.17)$$

Let us make numerical estimations of $\lambda$ and $\lambda^*$ taking into account the following typical molecular parameters: $a = 10^{-8}$ cm, $\mu = 10^{-23}$ g, $m = 10^{-22}$ g, $d = 10^{-18}$ e.s.u., $\theta = 10^{-26}$ e.s.u., $T = 300$ K and $N = 10^{17}/cm^3$. The frequencies $\omega$ will written as $\omega = A\, 10^{13}$ rad/s. With these values we verify that $\lambda$ and $\lambda^*$, defined, respectively, by Eqs. (3.16) and (3.17) are is given by

$$\lambda = 5.03\, 10^{15} \exp(-6.35\, A) \quad y^{-1}$$
and $\qquad\qquad\qquad\qquad\qquad\qquad\qquad\qquad\qquad\qquad\qquad (3.18),$
$$\lambda^* = 2.90\, 10^{12} \exp(-7.14\, A) \quad y^{-3/4}$$

measuring the time t in years (y).

Considering, for instance, $\omega = 4.3 \; 10^{13}$ rad/s (infrared region) and using Eqs.(1.3) and (3.17) we obtain $\tau = h/\delta \approx 3.37$ hours, $\lambda \approx 6969 \; y^{-1}$ and $\lambda^* \approx 0.13 \; y^{-3/4}$. Substituting these values into Eqs.(3.13) and (3.15), the *optical activity* for dilute gases $O_1(t)$ and for the dense gases and liquids $O_2(t)$ are given by

$$O_1(t) \approx \cos(t_1/3.37) \exp(-0.79 \; t_1)$$

and (3.19)

$$O_2(t) \approx \cos(t_1/3.37) \exp(-0.13 \; t_2^{3/4}),$$

where $t_1$ is measured in hours and $t_2$ in years.

According to Eqs.(3.19) the *optical activities* oscillate with a period $T_1 = 3.37$ hours around the zero average value and decays exponentially with time. For dilute gases $O_1(t) \to 0$ for $t_1 \gg 10$ hours and for dense gases or liquids $O_2(t) \to 0$ for $t \gg 50$ y. These results show that for $\omega = 4.3 \; 10^{13}$ rad/s the binary collisions in dilute gases are a very efficient racemization mechanism but very inefficient in dense gases and liquids.

**(4) $\varepsilon \neq 0$.**
(4.1) *U = static $\neq 0$.*

Putting $U = U(x)$ and $\varepsilon \neq 0$ into Eqs.(2.3) we verify that they can be exactly solved giving:[24]

$$a_L(t) = (a/\sqrt{2}) \cos \chi \; \exp(-iE_1t/\hbar) + (b/\sqrt{2}) \sin \chi \; \exp(-iE_2t/\hbar)$$

(4.1)

$$a_R(t) = -(a/\sqrt{2}) \sin \chi \; \exp(-iE_1t/\hbar) + (b/\sqrt{2}) \cos \chi \; \exp(-iE_2t/\hbar),$$

where a and b are constants determined by the initial conditions, $E_1 = E - \Delta$, $E_2 = E + \Delta$, $E = E_o + u$, $\cot 2\chi = \varepsilon/(\delta + \varphi)$ and $\Delta = [\varepsilon^2 + (\delta + \varphi)^2]^{1/2}$.

If at $t = 0$ the active molecule is prepared at the left configuration, that is, $|\Psi(0)\rangle = |L\rangle$, we obtain using Eqs.(4.1): $a = \sqrt{2} \cos \chi$ and $b = \sqrt{2} \sin \chi$. Therefore, the state $|\Psi(t)\rangle$ of the active molecular will described by:

$$|\Psi(t)\rangle = \exp(iEt/\hbar)\{[\cos^2\chi + \sin^2\chi \; \exp(2i\Delta t/\hbar)] \; |L\rangle$$
$$- i \sin(2\chi)\sin(\Delta t/\hbar) \; |R\rangle\} \quad (4.2)$$

Since $\sin(2\chi) = (\delta + \varphi)/[\varepsilon^2 + (\delta + \varphi)^2]^{1/2}$, the *racemization* r(t) will be given by

$$r(t) = |<R|\Psi(t)>|^2 = \Theta \sin^2([\varepsilon^2 + (\delta + \varphi)^2]^{1/2} t/\hbar) = \Theta[1 - \cos(2\Phi t/\hbar)]/2 \quad (4.3),$$

where the phase $\Phi = [\varepsilon^2 + (\delta + \varphi)^2]^{1/2}$ and the *racemization amplitude* $\Theta$ is given by $\Theta = (\delta + \varphi)^2/[\varepsilon^2 + (\delta + \varphi)^2]$. Consequently, the *optical activity* or *chiral activity* O(t) becomes

$$O(t) = 1 - \Theta + \Theta \cos(2\Phi t/\hbar) \quad (4.4).$$

From Eqs.(4.3)-(4.4) we verify that *chiral stability* or *optical stability* $O = 1$ can be achieved if the amplitude $\Theta \to 0$ that occurs only when the condition $\varepsilon \gg (\delta + \varphi)$ is obeyed.

Let us estimate the *racemization amplitude* $\Theta = (\delta + \varphi)^2/[\varepsilon^2 + (\delta + \varphi)^2]$, taking into account the parameters $\delta/h$ and $\varphi/h$ defined by Eqs.(1.3) and (3.10), respectively. In Fig.(2) is shown $\Theta = \Theta(A)$ as a function of A defined by the relation $\omega = A\, 10^{13}$ rad/s. In Fig.2 $\Theta(A)$ is plotted for two different $\varepsilon$ values: $\varepsilon/h = 10^{-3}$ Hz and $\varepsilon/h = 10^{-6}$ Hz.

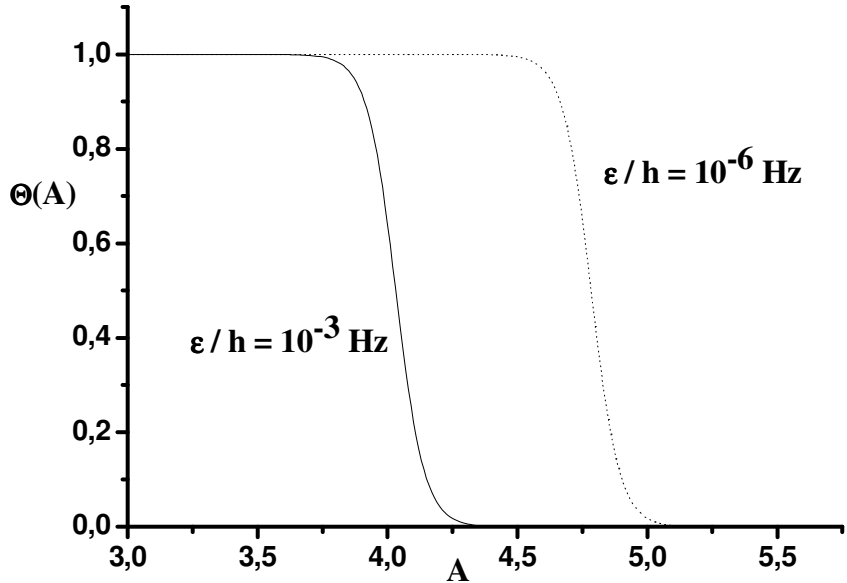

Figure 2. The *racemization amplitude* $\Theta(A)$, defined by Eq.(4.4), plotted as a function of the parameter A, defined by the equation $\omega = A\, 10^{13}$ rad/s. Two particular cases have been considered: $\varepsilon/h = 10^{-3}$ Hz (vibrational and rotational transitions) and $\varepsilon/h = 10^{-6}$ Hz (nuclear magnetic transitions).

From Fig.2 we see that in the case of rotational and vibrational transitions ($\varepsilon/h = 10^{-3}$ Hz) the *racemization amplitude* $\Theta(A) \to 0$ for frequencies $\omega > 4.25 \; 10^{13}$ rad/s. Thus, according to Eq.(4.4), for these frequencies the *optical activity* $O \to 1$ and, consequently, the sample can be *optically stable*. Similarly, when nuclear magnetic transitions ($\varepsilon/h = 10^{-6}$ Hz) are involved the *optical stability* or *chiral stability* is achieved, that is, $O \to 1$ only for $\omega > 5 \; 10^{13}$ rad/s since for these frequencies $\Theta \to 0$.

(4.2) *$U = 0$.*

When the active molecule is isolated $U = \varphi = 0$. In this case the amplitude $\Theta$ defined by Eq.(4.3) becomes

$$\Theta = \delta^2/(\varepsilon^2 + \delta^2) \qquad (4.5),$$

showing that the molecule can be *optically stable*, that is, $O \to 1$ only when $\varepsilon \gg \delta$. This last condition, as seen from Fig.(2), will depend on the nature of the molecular transition: rotational, vibrational or nuclear magnetic. The blocking effect of the weak interactions in the L–R oscillations, which occurs when $\varepsilon \gg \delta$, can be explained using the energy uncertainty relation $\Delta E \; \Delta t \geq h$. Indeed, since the spontaneous oscillation time $\tau = h/\delta$ between the L and R configurations, putting $\Delta t \sim \tau$ the energy uncertainty is given by $\Delta E \geq \delta$. In this way, if there is a difference of energy $\varepsilon$ between L and R, the natural L–R transitions are allowed only when $\Delta E \sim \delta \geq \varepsilon$. On the other side, the transitions will be prohibited when $\varepsilon \gg \delta$. In the presence of the static potential $\varphi$, using the same above reasoning, the L–R transitions are blocked when the condition $\varepsilon \gg \delta + \varphi$ is obeyed (see Section 4.1).

(4.3) *$U(t)$ produced by binary, additive and independent random collisions.*

In Section (3.3) we have calculated the *racemization* and the *optical activity* assuming that $\varepsilon = 0$ and that the active molecule is submitted to a time dependent potential $U(t)$ due to binary, additive and independent random collisions.

Now we will study the case when $\varepsilon \neq 0$ and $U(t)$ is generated by binary, additive and independent random collisions. Unfortunately, Eqs.(2.3) cannot be solved exactly when $\varepsilon \neq 0$ and $U = U(t)$. However, according to our precedent works[22–30] in these conditions the *racemization* $r(t)$ is given by

$$r(t) \approx (\delta/\Delta_o)^2 \; \{1 - \cos(2\Delta_o t/\hbar) \exp[-f(t)]\}/2 , \qquad (4.6)$$

where $\Delta_o = (\varepsilon^2 + \delta^2)^{1/2}$, $f(t) = \lambda t$ for dilute gases and $f(t) = \lambda^* t^{3/p}$ for compressed gases and liquids. The parameters $\lambda$ and $\lambda^*$ are defined by Eqs.(3.12) and (3.14), respectively. Using (4.6) the *optical activity* becomes,

$$O(t) = 1 - (\delta/\Delta_o)^2 + (\delta/\Delta_o)^2 \cos(2\Delta_o t/\hbar) \exp[-f(t)] \qquad (4.7).$$

From Eq.(4.7) we verify that for times t such that $\lambda t \gg 1$ or $\lambda^* t^{3/p} \gg 1$, O(t) tends asymptotically to $O(\infty) \to 1 - (\delta/\Delta_o)^2$. So, *chiral stability* can be achieved, that is, $O(\infty) \to 1$ if $\Delta_o \gg \delta$, that is, when $\varepsilon \gg \delta$. Using the parameter $\delta/h$ defined by Eq.(1.3) we verify that to get $O(\infty) \to 1$ the condition $\varepsilon/h \gg 5.54 \times 10^{12} A^{3/2} \exp(-9.52 A)$ Hz must be obeyed. So, for rotational and vibration transitions when $\varepsilon/h = 10^{-3}$ Hz we can easily verify that is occurs for A > 4.3, that is, for frequencies $\omega > 4.3 \times 10^{13}$ rad/s. Similarly, for nuclear magnetic transitions, when $\varepsilon/h = 10^{-6}$ Hz, the *chiral stability* is achieved only for frequencies $\omega > 5.4 \times 10^{13}$ rad/s.

## (5) Summary and conclusions.

We have analyzed, in the framework of the Schrödinger equation, the effect of intermolecular interactions on the tunneling racemization of the active molecule. The optically active molecule is assumed as a two-level system and the L–R isomerism was viewed in terms of a double-bottomed harmonic potential well. The active molecule is assumed to be embedded in a gas, liquid or solid, submitted to a perturbing potential U created by the molecules of the sample. In our model we have taken into account the difference of energy $\varepsilon$ due to the weak interactions between the left (L) and right (R) configurations.

When $\varepsilon = 0$ it was shown that the system cannot be *optically stable*. That is, the *optical activity* of the system, (1) oscillates periodically around zero when the molecules are isolated or submitted to a static potential and (2) tends asymptotically to zero in the case of dilute gases or compressed gases and liquids. The oscillation times (see Eqs.(3.6) and (3.8)) and the relaxation times (see Eqs.(3.13) and (3.15)) can be very large depending on the values of the parameters $\delta$, $\varphi$, $\lambda$ and $\lambda^*$ defined by the Eqs.(1.2),(3.9), (3.14) and (3.16), respectively.

When $\varepsilon \neq 0$ according to Eqs.(4.4),(4.5) and (4.7) the system can be *optically stable* only when $\varepsilon \gg (\delta + \varphi)$ and $\varepsilon \gg \delta$.

## (6) Discussions.

In a recent approach proposed by Vardi[7] to study the chiral stability, the self-consistent field has two components: $U_{hom}$ and $U_{het}$ emanating from the homochiral and heterochiral interactions, respectively. These components have been introduced in a nonlinear Schrödinger equation in order to give the time evolution of the active system. They have shown that when $U_{hom}$ interactions are energetically favorable to $U_{het}$ interactions, spontaneous L−R symmetry breaking may amplify the optical activity of a nearly racemic mixture.

Nonlinear quantum mechanics have been used[7,36] to explain the chiral stability. This seems to be a plausible attempt because the stationary states of a nonlinear Schrödinger`s equation[36,37] need not to be eigenstates of the operators that correspond to the symmetry group of the potential. So, the nonlinear term introduces a spontaneous symmetry breaking[37,38] which favors the localization in one of the wells. However, realistic nonlinear Schrödinger`s equations must be deduced taking into account exactly cooperative effects in the many-body interactions in the sample.[37-40] This algorithm would permit us to obtain a faithful nonlinear equation to study the optical stability. The nonlinear equations adopted by Vardi[7] and Koschany et al.[36] have not been obtained in this way. They have proposed, somewhat arbitrarily, equations following generic nonlinear models adopted in the literature.[41] In addition, we know that nonlinear equations exhibit a large number of rich and complex solutions depending on the magnitude of the nonlinear parameters. So, from the analysis of Vardi[7] and Koschany et al.[36], it is difficult to conclude that the nonlinear effects are, or are not, effective mechanisms responsible for the chiral stability.

Finally, it is important to remark that our conclusions regarding the stabilization of enantiomers are limited to those molecules that racemize only through L−R inversions. As is well known, there are many other different racemization mechanisms.[42] In our works these processes have not been considered.

**APPENDIX. Calculation of the static potential U = U(x).**

Let us assume that the active molecule is embedded in a dense gas, liquid or solid, where multiple interactions dominate over binary interactions and that there is a cooperative effect between the interacting molecules. Due to this collective behavior it will be assumed that each molecule is subjected to a *mean field* resulting from these combined interactions of all other molecules in the system. This *mean field* is understood as a *self-consistent Hartree field*.[33]

Let us consider the particular case of dense gases and liquids composed by dipolar molecules. This is a special case because a cooperative interaction mechanism appears between the molecules of the sample and U(x) can be easily calculated. To do this we assume, in a first approximation, that the active molecule is inside a small cavity, with radius R, surrounded by the perturbing ones. According to Claverie and Jona-Lasinio[33], once the active molecule is in a localized configuration, │L > or │R>, it has a non-zero average dipole moment **d** = < **d** >, then this moment locally polarizes the surrounding which, in turn creates, at the position of **d**, a so-called *reaction field* $\mathbf{E_r}$, which is collinear with **d**. As the interaction −**d**·$\mathbf{E_r}$ is negative it tends to stabilize the non-symmetric state under consideration. The reaction field, which is clearly a nonlinear effect,[33] is the statistical mechanics average < **E** > of the electric field **E** created by the molecules surrounding the dipole **d**. It is a standard topic in the theory of the dielectric constant and of the solvent effects.[33–35] The reaction field $\mathbf{E_r}$ corresponding to a dipole **d** embedded in a spherical cavity of radius $R$ inside a medium with dielectric constant $\varepsilon$ is given[33] by $\mathbf{E_r}=2(\varepsilon − 1)\mathbf{d}/[(2\varepsilon + 1)R^3]$. In this way the interaction potential U(x) (between this electric field and the active molecule) is given by U(x) = − **d** $\mathbf{E_r}$. Since the dipole matrix element of the active molecule between │L > and │R > states is zero, the heterochiral interaction < L│U(x)│R> of this molecule with $\mathbf{E_r}$ will be calculated taking into account the quadrupole moment Q(x) of the active molecule. So, < L│U(x)│R> = φ will be given by φ ≈ d <L│Q(x)│R>/$R^4$. We have shown elsewhere[23] that <L│Q(x)│R> is given by <L│Q(x)│R> = θ exp(−μωa²/ℏ), where θ is the quadrupole matrix element of the active molecule between left and right configurations. So, φ is given by

$$\varphi = (\theta d/R^4) \exp(-\mu\omega a^2/\hbar) \qquad (A.1).$$